\documentclass{elsart5}

 \usepackage{graphicx}

\usepackage{amssymb}

\begin{document}


\title{Checkerboard order in the t--J model on the square lattice}

\author[aff1,aff3,aff4]{D.~Poilblanc\corauthref{cor1}}
\ead{didier.poilblanc@irsamc.ups-tlse.fr}
\ead[url]{http://www.lpt.ups-tlse.fr/}
\corauth[cor1]{}
\author[aff2]{C.~Weber}
\author[aff3]{F.~Mila}
\author[aff4]{M.~Sigrist}
\address[aff1]{ Laboratoire de Physique Th\'eorique UMR 5152, C.N.R.S. \&
Universit\'e de Toulouse, F-31062 Toulouse, France}
\address[aff2]{Institut Romand de Recherche Num\'erique en 
Physique des Mat\'eriaux (IRRMA),
  PPH-Ecublens, CH-1015 Lausanne, Switzerland}
\address[aff3]{Institute of Theoretical Physics,
Ecole Polytechnique F\'ed\'erale de Lausanne,
BSP 720,
CH-1015 Lausanne,
Switzerland}
\address[aff4]{Theoretische Physik, ETH-H\"onggerberg, 8093 Z\"urich,
Switzerland}
\received{12 June 2005}
\revised{13 June 2005}
\accepted{14 June 2005}


\begin{abstract}
We propose that the inhomogeneous patterns seen by STM in
some underdoped superconducting cuprates could be related
to a bond-order-wave instability of the staggered flux state, one of the 
most studied "normal" state proposed to compete with the d-wave RVB
superconductor. A checkerboard pattern is obtained
by a Gutzwiller renormalized mean-field theory of the t-J
model for doping around 1/8. It is found that the charge modulation 
is always an order of
magnitude smaller than the bond-order modulations. This is confirmed by
an exact optimization of the wavefunction by a variational Monte Carlo
scheme. The numerical estimates of the order parameters 
are however found to be strongly reduced
w.r.t their mean-field values.

\end{abstract}



\section{Introduction}\label{Introduction}

With  constantly improving resolution of experimental
techniques, novel features in
the global phase diagram of high-T$_c$ cuprate superconductors
have emerged. One of the most striking is the observation by
scanning tunnelling
microscopy (STM), in the pseudogap phase of underdoped
Bi$_2$Sr$_2$CaCu$_2$O$_{8+\delta}$~\cite{STM-BSCO}  and 
Ca$_{2-x}$Na$_x$CuO$_2$Cl$_2$ single
crystals~\cite{STM_CaNaCOCl}, of a form of local electronic ordering, 
with a spatial period close to four lattice spacings. 
These observations raise important theoretical
questions about the relevance of such structures in the framework
of strongly correlated models.
Here, we analyze the stability of new inhomogeneous 
phases~\cite{Inhomog_DP,Inhomog_Li,Inhomog_Weber}
which can compete in certain conditions with the d-wave superconducting RVB
state~\cite{RVB}. Both mean-field~\cite{Inhomog_DP} (MF) and numerical
Variational Monte Carlo results~\cite{Inhomog_Weber}(VMC) 
will be summarized here.

\section{Renormalized Mean-field Theory}

We describe the doped antiferromagnet
by a $t{-}J$ model,
  \begin{equation}
   H= - t\sum_{\langle ij\rangle\sigma}
      (c^{\dagger}_{i,\sigma}c_{j,\sigma}+h.c.)
    + J\sum_{\langle ij\rangle} {\bf S}_i \cdot {\bf S}_j \, .
  \end{equation}
First, we replace the local constraints of no doubly occupied sites
 by statistical
Gutzwiller weights (see below) and use a mean-field 
decoupling in the particle-hole channel to obtain a self-consistent 
renormalized MF hamiltonian~\cite{Renormalized_MF},
  \begin{eqnarray}
   H_{\rm MF}= &-& t \sum_{\langle ij\rangle\sigma} g_{ij}^t
      (c^{\dagger}_{i,\sigma}c_{j,\sigma}+h.c.)\nonumber \\
   &-&\frac{3}{4} J \sum_{\langle ij\rangle\sigma}g_{ij}^J
(\chi_{ji}c^{\dagger}_{i,\sigma}c_{j,\sigma} + h.c. -|\chi_{ij}|^2)\, ,
  \end{eqnarray}
where $\chi_{ji}=\langle c^\dagger_{j,\sigma}c_{i,\sigma}\rangle$
and where we explicitly assume an inhomogeneous solution and hence
inhomogeneous Gutzwiller factors,
  \begin{eqnarray}
\label{Eq:Gutz}
   g_{ij}^J&=\frac{4(1-x_i)(1-x_j)}{(1-x_i^2)(1-x_j^2)-8x_ix_j|\chi_{ij}|^2
+16|\chi_{ij}|^4}\, ,\\
  g_{ij}^t&=[\frac{4x_ix_j(1-x_i)(1-x_j)}
{(1-x_i^2)(1-x_j^2)+8(1-x_ix_j)|\chi_{ij}|^2
+16|\chi_{ij}|^4}]^{\frac{1}{2}}\, ,\nonumber 
  \end{eqnarray}
where $x_i$, $x_j$ are the local hole densities.
The $|\chi_{ij}|$ terms account for the correlations
of the probabilities between nearest-neighbor
sites~\cite{Renormalized_MF_2}
and were not included in the
original MF treatment~\cite{Inhomog_DP} of the inhomogeneous 
state.
Our MF scheme applies to the "normal" phase. In contrast, 
an additional decoupling in the particle-particle channel leads to
an RVB superconducting state~\cite{RVB}.

Interestingly, the parent insulating RVB state
can be viewed as a staggered 
flux state (SFP)~\cite{half-flux,kotliar_fluxphases,SFP_bis} MF
solution of the above hamiltonian. We show here that, away from half-filling,
the SFP spontaneously acquires (small) bond modulations. 

\section{Results and Comparison with VMC}

\begin{figure}[htb]
\centerline{\includegraphics[width =0.9\columnwidth]{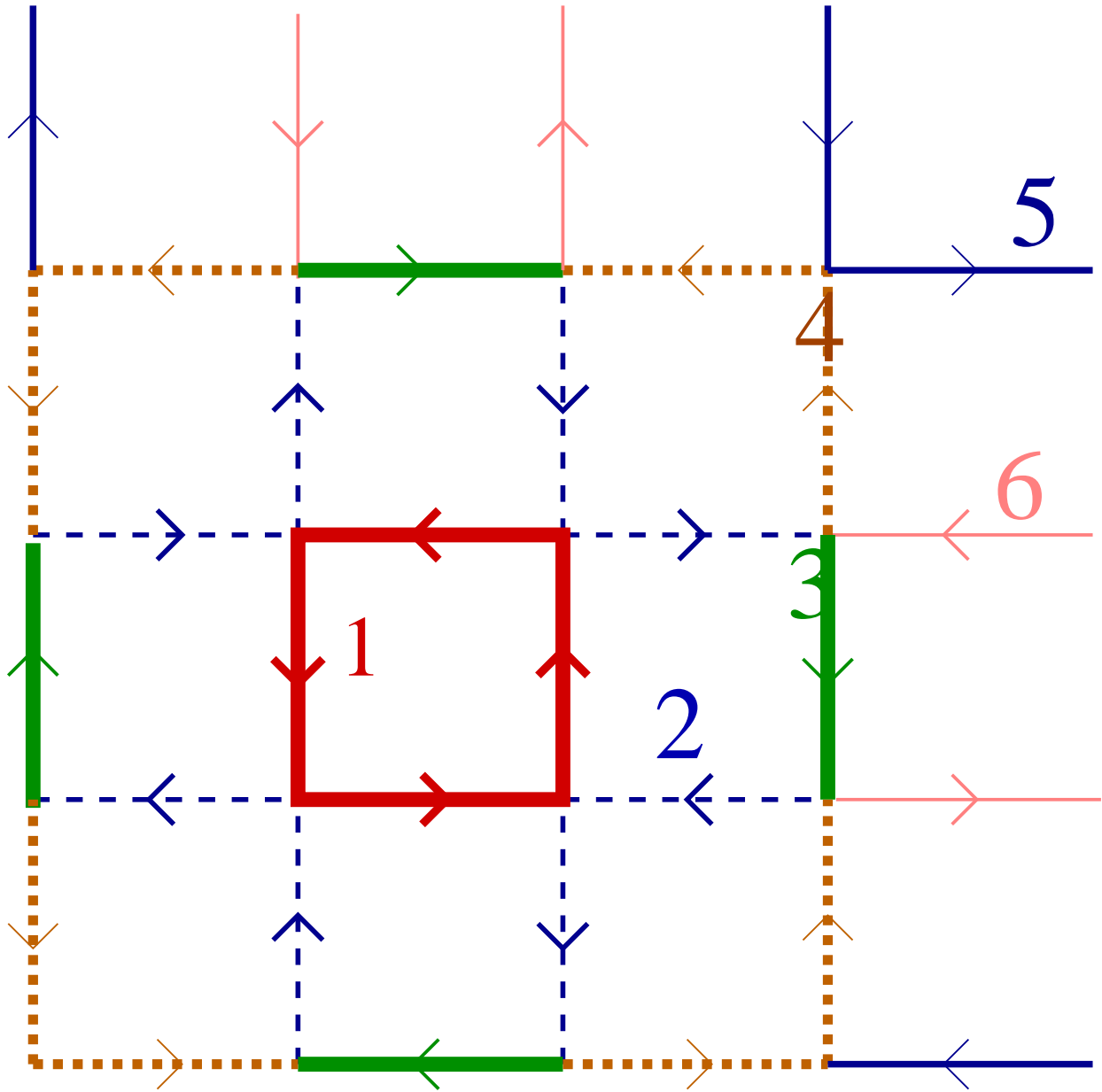}}
\caption{Schematic pattern of the bond
distribution. Non-equivalent bonds (labelled from 1 to 6) are
indicated by different types of lines whose widths qualitatively
reflect the bond magnitudes. Arrows indicate the
directions of the small charge currents. 
}
  	\label{fig-1}
  \end{figure}

Assuming a $4\times 4$ supercell with overall rotational symmetry 
(see Fig.~\ref{fig-1}), the MF equations 
have been solved for an average doping of 1/8.
A typical solution depicted in Fig.~\ref{fig-1} exhibits orbital currents
according to the original SFP pattern. However,
small modulations of the bond hoppings 
$\big<c^{\dagger}_{i,\sigma}c_{j,\sigma}\big>$ and bond exchange
terms $\big<{\bf S}_i \cdot {\bf S}_j\big>$ as well as the on-site densities
appear spontaneously. Numerical values of
the bond (site) parameters for $t/J=3$ are given in
Table~\ref{table:bond} (Table~\ref{table:site}).
The charge-density-wave (CDW) modulation 
is found to be an order of magnitude smaller than the bond-order-wave (BOW)
modulations. It was argued recently~\cite{Inhomog_Li} that this instability 
is connected to nesting properties of the small ellipsoidal 
Fermi pockets of the SFP. 

The MF results have been tested by an exact (within statistical
errors) VMC energy minimization and evaluation of the order
parameters (on a $16\times 16$ lattice with mixed boundary
conditions,  see Ref.~\cite{Inhomog_Weber} for details).
Comparative values are shown in 
Tables~\ref{table:bond} \& \ref{table:site}. 
Interestingly, the VMC simulation also predicts a dominant BOW 
component w.r.t the CDW one.
However, an overall reduction of the magnitudes
of these modulations by an
approximate factor of 4 compared to MF is seen.

How such a state can compete with the d-wave RVB superconductor is
still unclear. However, our
calculations~\cite{Inhomog_DP,Inhomog_Weber} 
have suggested that the BOW SFP is less sensitive than the
superconducting state to a moderate
(e.g. nearest-neighbor) Coulomb repulsion. In this competition, the
role of the disorder and/or the surface might be crucial.  

\begin{table}[htb]
\caption{\label{table:bond}
Numerical values of the spin-spin correlations (first 3 lines) and
of the real (next 3 lines) and imaginary (last 3
lines) parts of the bond hoppings 
$\big<\sum_\sigma c_{i\sigma}^\dagger c_{j_\sigma}\big>$ 
in the BOW stagered flux state
for $t/J=3$ and $x\simeq 1/8$ 
(see Table~\protect\ref{table:site} for exact densities). 
(1) from Ref.~\protect\cite{Inhomog_DP}; (2) with corrections in
Gutzwiller factors. 
Two last columns: average and relative BOW modulations (in $\%$).}
\begin{center}
\begin{tabular}{|l|l|l|l|l|l|l|l|l|}
\hline
Bond $\#$ & 1& 2& 3&   4&   5& 6 & {\rm aver.} & $\%$ ($\pm$)\\
\hline
MF (1)
& -0.346 & -0.172 & -0.313 & -0.236 &  -0.164 & -0.125 & -0.220& 50$\%$\\
MF (2)
& -0.258 & -0.221 & -0.310 & -0.194 &  -0.170 & -0.184 & -0.219& $32\%$\\
{\bf VMC} 
&-0.2149  & -0.2062 & -0.2128 & -0.1871  & -0.1810   & -0.1750 &
-0.1963& {\bf 10$\%$}\\
&   &   & & & & & & \\
MF (1)
& 0.055 & 0.057 & 0.119 & 0.106 & 0.092 & 0.073& 0.083 & 38$\%$\\
MF (2)
& 0.057 & 0.054 & 0.061 & 0.081 &  0.112 & 0.051 & 0.069 & 44$\%$\\
{\bf VMC} 
& 0.0706 & 0.0742  & 0.0840 & 0.0833 & 0.0855  & 0.0825  & 0.0797&
{\bf 9.5$\%$}\\
&   &   &  & & & & & \\
MF (1)
& 0.045 & 0.037 & 0.037 & 0.031 & 0.023 & 0.030& 0.034 & $32\%$\\
MF (2)
& 0.038 & 0.038 & 0.044 & 0.031 &  0.011 & 0.036& 0.033& $50\%$\\
{\bf VMC} 
& 0.0400 & 0.0392  &0.0414  & 0.0349 & 0.0313 & 0.0330  & 0.0367&
 {\bf 12$\%$}\\
\hline
\end{tabular}
\end{center}
\end{table}

\begin{table}[htb]
\caption{\label{table:site} Numerical values of the charge densities
on the 3 non-equivalent sites
(labelled from the center of the $4\times 4$ pattern) 
for $t/J=3$ and doping $x\simeq 1/8$.
Average density and relative amplitudes (in $\%$) are 
provided in the last column.}
\begin{center}
\begin{tabular}{|l|l|l|l|l|l|}
\hline
Site $\#$ & 1& 2& 3 & $n$ & $\%$ ($\pm$)\\
\hline
MF (1) & 0.925& 0.864 & 0.857 & 0.8776 & 4$\%$\\
MF (2) & 0.901& 0.898 & 0.806 & 0.8759 & 5.5$\%$\\
 {\bf VMC} & 0.886(3) & 0.874(5) & 0.863(6)  & 0.875 &  {\bf 1.3$\%$}\\
\hline
\end{tabular}
\end{center}
\end{table}


\begin{thebibliography}{00}

\bibitem{STM-BSCO} M.~Vershinin, S.~Misra, S.~Ono, Y.~Abe,
Y.~Ando, and A.~Yazdani, Science {\bf 303}, 1995 (2004). 



\bibitem{STM_CaNaCOCl} T.~Hanaguri et al., Nature {\bf 430}, 1001 (2004).

\bibitem{Inhomog_DP} D.~Poilblanc, Phys.~Rev.~B. 
{\bf 72}, 060508(R) (2005)

\bibitem{Inhomog_Li} C.~Li, S.~Zhou and Z.~Wang, Phys.~Rev.~B. 
{\bf 73}, 060501(R) (2006); for related results based on an
instability
critirium see also, Z.~Wang, G.~Kotliar and X.F.~Wang, Phys.~Rev.~B.
{\bf 42}, 8690 (1990).

\bibitem{Inhomog_Weber} C.~Weber, D.~Poilblanc, S.~Capponi, F.~Mila
  and C.~Jaudet, cond-mat/0601301.

\bibitem{RVB} P.W.~Anderson, Science {\bf 235},
1196 (1987).

\bibitem{Renormalized_MF} F.C.~Zhang, C.~Gros, T.M.~Rice and H.~Shiba,
Supercond.~Sci.~Technol.~{\bf 1}, 36 (1988).

\bibitem{Renormalized_MF_2} T.C.~Hsu, Phys.~Rev.~B {\bf 41} 11379 (1990);
M.~Sigrist, T.M.~Rice and F.C.~Zhang, Phys.~Rev.~B {\bf 49} 12058 (1994).

\bibitem{half-flux} I.~Affleck and J.B.~Marston,
Phys.~Rev.~B. {\bf 37}, R3774 (1988); J.B.~Marston and I.~Affleck,
{\it ibid.} {\bf 39}, 11538 (1989).

\bibitem{kotliar_fluxphases} G.~Kotliar, Phys.~Rev.~B {\bf 37},
3664 (1988). 

\bibitem{SFP_bis}
D.~Poilblanc and Y.~Hasegawa, Phys.~Rev.~B {\bf
41}, 6989 (1990);
M.U.~Ubbens and P.A.~Lee, Phys.~Rev.~B {\bf 46}, 8434 (1992);
P.A.~Lee,  N.~Nagaosa, T.~Ng and X.~Wen , Phys.~Rev.~B
{\bf 57}, 6003 (1998);
D.A.~Ivanov, Phys.~Rev.~B. {\bf 70}, 104503 (2004).

\end{thebibliography}
\end{document}